\documentclass{article}
\usepackage[margin=1.2in]{geometry}

\usepackage{amsmath,amssymb,amsfonts,graphicx,amsthm,nicefrac,mathtools,bm}
\usepackage{hyperref}
\usepackage[numbers]{natbib}
\usepackage{bbm}
\usepackage{color}
\usepackage{xspace}
\newtheorem{theorem}{Theorem}

\theoremstyle{definition}

\theoremstyle{remark}

\makeatletter
\newtheorem*{rep@theorem}{\rep@title}
\newcommand{\newreptheorem}[2]{%
\newenvironment{rep#1}[1]{%
\def\rep@title{#2 \ref{##1}}%
\begin{rep@theorem}}%
{\end{rep@theorem}}}
\makeatother
\newreptheorem{theorem}{Theorem}
\newreptheorem{lemma}{Lemma}
\newreptheorem{corollary}{Corollary}
\newreptheorem{proposition}{Proposition}

\DeclareMathOperator{\sign}{sign}

\DeclareMathOperator*{\argmin}{arg\,min}

\newcommand{\iid}[0]{i.i.d.\xspace}

\newcommand{\norm}[1]{\lVert{#1}\rVert}

\newcommand{\PP}[1]{\mathbb{P}\left\{{#1}\right\}} 

\newcommand{\EEst}[2]{\mathbb{E}\left[{#1}\ \middle| \ {#2}\right]} 

\def\R{\mathbb{R}}

\newcommand{\ident}{\mathbf{I}}

\newcommand{\iidsim}{\stackrel{\mathrm{iid}}{\sim}}

\usepackage{fancyhdr}
\newcommand{\thedate}{\today}
\newcommand{\theauthor}{}
\newcommand{\thetitle}{}

\date{\thedate}
\author{\theauthor}
\title{\thetitle}

\fancypagestyle{plain}{%
\fancyhf{}%
\lhead{\thepage}
}

\usepackage[mmddyy]{datetime}

\newcommand{\PI}{\textnormal{PI}}

\newcommand{\PIr}{\PI_{\textnormal{rounded}}}

\newcommand{\Yhat}{\widehat{\mathcal{Y}}}
\newcommand{\Yt}{\widetilde{\mathcal{Y}}}
\newcommand{\dhat}{\widehat{d}}
\newcommand{\dt}{\widetilde{d}}
\newcommand{\Alg}{\mathcal{A}}
\newcommand{\muhat}{\widehat{\mu}}
\newcommand{\mut}{\widetilde{\mu}}
\newcommand{\tR}{\widetilde{R}}

\title{Discretized conformal prediction for efficient distribution-free inference}
\author{Wenyu Chen, Kelli-Jean Chun, and Rina Foygel Barber}
\date{Sept.~2017 (updated Jan.~2023)}

\begin{document}
\maketitle

\begin{abstract}
In regression problems where there is no known true underlying model,
conformal prediction methods enable prediction intervals to be constructed
without any assumptions on the distribution of the underlying data,
except that the training and test data are assumed to be exchangeable.
However, these methods bear a heavy computational cost---and, to be carried
out exactly, the regression algorithm would need to be fitted infinitely
many times. In practice, the conformal prediction method is 
run by simply considering only a finite grid of finely spaced values for the response variable. 
This paper develops discretized conformal prediction algorithms
that are guaranteed to cover the target value with the desired probability,
and that offer a tradeoff between computational cost and 
prediction accuracy.
\end{abstract}

\section{Introduction}
In this paper, we examine the problem of efficiently computing conformal prediction intervals using models that are computationally expensive to fit.
The conformal prediction method, introduced by \citet{vovk1999,vovk2005,vovk2009} and developed for the high-dimensional setting by \citet{lei2016}, uses a training data set $(X_1,Y_1),\dots,(X_n,Y_n)\in\R^p\times \R$ to provide a prediction interval for an unobserved response variable $Y_{n+1}\in\R$ at the covariate point $X_{n+1}\in\R^p$. 
The prediction interval's coverage guarantees rely  only on the assumption that the available training data $(X_1,Y_1),\dots,(X_n,Y_n)$ is exchangeable with the test data point $(X_{n+1},Y_{n+1})$.

As originally proposed, this method requires refitting an expensive model for every possible value of the test point's response variable $Y_{n+1}$---at least in theory, but of course in practice, if $Y_{n+1}$ is real-valued, it is impossible to refit the model infinitely many times, and so instead it is common to round $Y_{n+1}$ to a fine finite grid of values in $\R$.

Our work formalizes this rounding procedure, proving that rounding can be done without losing the coverage guarantee of the method. Our result also allows for the rounding to be as coarse or fine as desired, giving a principled way to trade off between computational cost and the precision of the prediction (as measured by the width of the prediction interval), while maintaining the coverage guarantee. 

\section{Background}

Given a training data set $(X_1,Y_1),\dots,(X_n,Y_n)\in\R^p\times \R$, and a new feature 
vector $X_{n+1}\in\R^p$, the goal of predictive inference is
to provide an
 interval in $\R$ that is likely to contain the unobserved response value $Y_{n+1}$.
Imagine fitting a predictive model $\muhat:\R^p\rightarrow\R$, where $\mu(x)$
predicts the value of $Y$ given $X=x$, to the $n$ training points.
If $Y_i$ is within the interval $\muhat(X_i)\pm \epsilon$ for $90\%$ of the training
data points $i=1,\dots,n$, we might naively assume that $\muhat(X_{n+1})\pm \epsilon$
is a $90\%$-coverage prediction interval for $Y_{n+1}$, that is, that $\PP{Y_{n+1}\in\muhat(X_i)\pm\epsilon} = 0.9$. However, for high dimensions $p$,
in general this will be completely untrue---the model $\muhat$, having been fitted
to the training data points, by its construction will have low residuals $|Y_i - \muhat(X_i)|$
on this same training data set, but may be wildly inaccurate on an independently
drawn test point $(X_{n+1},Y_{n+1})$. In general, the constructed prediction
interval $\muhat(X_{n+1})\pm \epsilon$ will {\em undercover}, i.e.~the probability
of this interval containing (``covering'') the true response value $Y_{n+1}$
will be lower than intended.

The problem is that while the training and test data ($n+1$ total data points)
may have been drawn from the same distribution initially, the resulting $n+1$
many residuals
are {\em not} exchangeably distributed since $\muhat$ was fitted on the $n$ training
points without including the test point.

At first glance, this problem seems insurmountable---without observing the test
point, how can we hope to include it into the process of fitting the model $\muhat$?
Remarkably, the conformal prediction method offers a way to do exactly this,
which can be 
carried out using any desired model fitting algorithm for constructing $\muhat$.
Here we summarize the steps of the conformal prediction method,
as presented in the work of~\citet{lei2016}.

\begin{itemize}
\item[(1)] Without looking at the data, we choose any model fitting algorithm
\[\Alg: \Big((x_1,y_1),\dots,(x_{n+1},y_{n+1})\Big)\mapsto \muhat\]
that is required to treat the $n+1$ many input points exchangeably but is otherwise unconstrained.
\item[(2)]
Given the data, we compute
\[\muhat_y = \Alg\Big((X_1,Y_1),\dots,(X_n,Y_n),(X_{n+1},y)\Big)\]
for every value $y\in\R$---each $y$ is a possible value for the unseen test data point's response value, $Y_{n+1}$.
\item[(3)] Compute the desired  quantile for the residuals,
\[Q_y = \textnormal{Quantile}_{(1-\alpha)(1+1/n)}\left\{\big|Y_i - \muhat_y(X_i)\big|:i=1,\dots,n\right\},\]
where $\alpha$ is the predefined desired error level.

\item[(4)] 
The  prediction interval\footnote{While the prediction set is labeled PI for ``prediction interval'', and we often refer to it with this terminology, in some settings the set might not be equal to a single interval.} is given by
\[\PI = \left\{y\in\R : y\in \muhat_y(X_{n+1})\pm Q_y\right\}.\]
\end{itemize}

The conformal prediction algorithm offers a coverage guarantee with no assumptions aside from exchangeability of the data points (for example, if the training and test points are i.i.d.~draws from some distribution).
\begin{theorem}[{\citet[Theorem 2.1]{lei2016}}]\label{thm:conformal}
Assume exchangeability of the training and test data points
\[(X_1,Y_1),\dots,(X_n,Y_n),(X_{n+1},Y_{n+1}).\]
Then the conformal prediction algorithm satisfies
\[\PP{Y_{n+1}\in\PI}\geq 1-\alpha.\]
\end{theorem}
We reproduce a short proof of this result
here,
as this proof technique will be useful
for proving the results presented later in this paper.
\begin{proof}[Proof of Theorem~\ref{thm:conformal}]
Define residuals
$R_i = Y_i - \muhat_{Y_{n+1}}(X_i)$
for each $i=1,\dots,n+1$. Since $\muhat_{Y_{n+1}}$ is a fitted model that was constructed using the $n+1$ many data points exchangeably, we therefore see that $R_1,\dots,R_n,R_{n+1}$ are themselves exchangeable, and so
\[\PP{|R_{n+1}|\leq \textnormal{Quantile}_{1-\alpha}\Big\{|R_i|:i=1,\dots,n+1\Big\}}\geq 1-\alpha.\]
By a simple calculation this event is equivalent to
\[\big|Y_{n+1} - \muhat_{Y_{n+1}}(X_{n+1})\big| = |R_{n+1}|\leq \textnormal{Quantile}_{(1-\alpha)(1+1/n)}\Big\{|R_i|:i=1,\dots,n\Big\} = Q_{Y_{n+1}},\]
where we use the definitions of $R_{n+1}$ and $Q_{Y_{n+1}}$. In other words, the bound
$|y - \muhat_y(X_{n+1})| \leq Q_y$
holds for $y=Y_{n+1}$. By definition, 
this means that $Y_{n+1}\in\PI$, proving
the theorem.\end{proof}

\paragraph{Computation for conformal prediction}
Examining the conformal prediction algorithm,
the reader may notice that for each 
possible value $y\in\R$ (that is, for
each potential $y$ value for the test 
data point, $Y_{n+1}$), we must
refit a model $\muhat_y$ using the 
algorithm $\Alg$.
Depending on the setting, each 
run of $\Alg$ may be fairly expensive---and
even disregarding cost, in general
we cannot hope to run $\Alg$ infinitely
many times, once for each $y\in\R$.

In some settings, this problem
can be circumvented using special structure
within the model fitting algorithm. For instance,
if $\Alg$ fits a linear model with  a squared $\ell_2$ norm penalty (Ridge regression), the prediction interval
$\PI$ in fact enjoys
a closed form solution~\cite{burnaev2014efficiency}.
Recent work by~\citet{lei2017}
provides an efficient method for computing
$\PI$ for the Lasso, i.e.~quadratic
loss function + $\ell_1$ norm penalty.

In nearly any other setting, however,
we must instead turn to approximations of
the full conformal prediction method,
since fully computing $\PI$ is impossible.
A straightforward way to approximate
the algorithm is to only fit $\muhat_y$ for a finite set of $y$ values---for instance,
taking a fine grid of $y$ values over some interval $[a,b]$ that includes the empirical range of the observed response values, $a\leq \min_{1\leq i\leq n}Y_i\leq \max_{1\leq i\leq n}Y_i\leq b$---we give more details below. 
This range may be further reduced for greater computational efficiency via ``trimming'', as in~\citet{chen2016trimmed}. 
An alternate approach is to employ sample splitting, studied by~\citet{lei2016}, where half the training data
is used to fit the model $\muhat$ a single time, while the quantiles of the residual
are then computed over the remaining $n/2$ data points. Split conformal prediction is highly efficient, requiring only a single run of the model fitting algorithm $\Alg$, but may produce substantially wider prediction intervals due to the effective sample size being reduced to half.
(Of course, it is also possible
to create a uneven split, using a larger portion of data for model fitting and a smaller set for the inference step. This will produce sharper prediction intervals, but the method will have higher variance; this tradeoff is unavoidable for data splitting methods.)

\subsection{Approximation via rounding}\label{sec:background_rounding}
As mentioned above, in most settings,
in practice the model $\muhat_y$
can only be fitted over some finite 
grid of $y$ values spanning some range $[a,b]$. Specifically, a common approximate algorithm might take the form:
\begin{itemize}
\item[(1)] Choose $\Alg$ as before, and a finite set $\Yhat=\{y_1,\dots,y_M\}$
of trial $y$ values, with spacing $\Delta = (b-a)/(M-1)$, i.e.~$y_m = a + (m-1)\Delta$.
\item[(2)] \hspace{-.08in},(3)
As before, Compute $\muhat_{y_m}$ and $Q_{y_m}$ for each trial $y$ value, i.e.~for $m=1,\dots,M.$
\item[(4)] 
The rounded prediction interval is given by
\[\PIr = \left\{m\in\{1,\dots,M\} : y_m\in \muhat_{y_m}(X_{n+1})\pm Q_{y_m}\right\}.\]
Then extend by a margin of $\Delta$ to each side:
\[\PI = \bigcup_{m\in\PIr} \big(y_m-\Delta,y_m+\Delta\big).\]
\end{itemize}

In practice, this type of approximate conformal prediction algorithm performs well, but there are several drawbacks. First, from a theoretical point of view, coverage can no longer be guaranteed---in particular, if there is some $y$ value that lies between two grid points, $y_m<y<y_{m+1}$, it is possible that neither $y_m$ nor $y_{m+1}$ gets selected 
by the discretized algorithm, but we would have placed $y$ itself into the prediction interval $\PI$ if $y$ had been one of the values tested. In general this can happen only if the true prediction interval (the $\PI$ from the original non-discretized method) does not consist of a single connected component---depending on the model fitting algorithm $\Alg$, this may or may not occur. Second, in practice, the spacing $\Delta$ of the grid provides a lower bound on the precision of the method---the set $\PI$ will always be at least $2\Delta$ wide. Since $\Delta\propto M^{-1}$, this forces us to use a large computational budget $M$.

We will next propose two different approaches towards a discretized conformal prediction method, which will resolve these issues by allowing for theoretical coverage properties and, in one of the algorithms, for prediction intervals whose width may be narrower than the spacing of the discretized grid.

\section{Main results}

In this section we introduce two different versions of discretized conformal inference, with a coverage guarantee for both algorithms given in Theorem~\ref{thm:rounding} below.

\subsection{Conformal prediction with discretized data}\label{sec:alg1}
We begin with a simple rounded algorithm for conformal prediction, where our analysis is carried out entirely on the rounded data---that is, all response values $Y_i$
are rounded to some finite grid---before converting back to the original
values as a final step.
\begin{itemize}
\item[(1)] 
Without looking at the data, choose any model fitting algorithm
\[\Alg: \Big((x_1,y_1),\dots,(x_{n+1},y_{n+1})\Big)\mapsto \muhat,\]
where $\muhat$ maps a vector of covariates $x$ to a predicted value for $y$ in $\R$.
The model fitting algorithm $\Alg$ is required to treat the $n+1$ many input points exchangeably but is otherwise unconstrained.
Furthermore, choose a set $\Yhat\subset\R$ containing finitely many points---this set is the ``grid'' of candidate values for the response variable $Y_{n+1}$ at the test point.
Select also a discretization function $\dhat:\R\rightarrow\Yhat$ that rounds response values $y$ to values in the grid $\Yhat$. 

\item[(2)] Next, apply conformal prediction to this rounded data set. Specifically, we compute
\[\muhat_y = \Alg\Big((X_1,\dhat(Y_1)),\dots,(X_n,\dhat(Y_n)),(X_{n+1},y)\Big)\]
for possible value $y\in\Yhat$.
\item[(3)] Compute the desired  quantile for the residuals,
\[Q_y = \textnormal{Quantile}_{(1-\alpha)(1+1/n)}\left\{\big|\dhat(Y_i) - \muhat_y(X_i)\big|:i=1,\dots,n\right\},\]
where $\alpha$ is the predefined desired error level.

\item[(4)] 
The discretized prediction interval is given by
\[\PIr = \left\{y\in\Yhat : y\in \muhat_y(X_{n+1})\pm Q_y\right\}.\]
This prediction interval is, by construction, likely to cover the {\em rounded} test response value, $\dhat(Y_{n+1})$.
 To invert the rounding step, the final prediction interval
is then given by
\[\PI = \dhat^{\,-1}(\PIr) = \{y\in\R: \dhat(y) \in \PIr\}.\]
\end{itemize}

The reason for the notation $\Yhat$, for the grid of candidate $y$ values, 
is that in practice the grid is generally
determined as a function of the data (and, therefore, the same may be true for the discretization function $\dhat$).
Most commonly, the grid might be determined by taking $m$ equally spaced points from some minimum value $y_{\min}$ to some maximum value $y_{\max}$, where $y_{\min},y_{\max}$ are determined by the empirical range of response values in the training data, i.e.~by the range of $Y_1,\dots,Y_n$. The function $\dhat$ would then simply round to the nearest value in this grid. (The number of points, $m$, is more commonly independent of the data, and simply depends on our computational budget---how many times we are willing to refit the model.)

To formalize the setting where $\Yhat$ and $\dhat$ depend on the data, we let 
\[\Yt = \Yt\Big((X_1,Y_1),\dots,(X_{n+1},Y_{n+1})\Big)\subset\R,\]
be any finite set, let
\[\dt = \dt\Big((X_1,Y_1),\dots,(X_{n+1},Y_{n+1})\Big)\in\{d:\R\rightarrow \Yt\}\]
be any function mapping to that set, that depend arbitrarily on the training and test data; however, $\Yt$ and $\dt$ are constrained to be exchangeable functions of the data.
If $\Yhat$ and $\dhat$ are nearly always equal to $\Yt$ and $\dt$---as is the case when $\Yhat$ depends only on the range of the $Y_i$'s, and $\dhat$ simply rounds to the nearest value---then the fact that $\Yhat$ and $\dhat$ depend on the data will only slightly affect coverage.

\subsection{A better way to round: conformal prediction with a discretized model}\label{sec:alg2}
While the naive rounded algorithm presented above, where the data
itself is discretized, will successfully provide the correct coverage guarantees, it may be overly conservative. In particular, the prediction intervals will always need to be at least as wide as the interval between two grid points (as was also the case with the rounding approximation presented in Section~\ref{sec:background_rounding}). We now modify our algorithm to more fully use the information in the data, and hopefully to attain narrower intervals.
Specifically, instead of discretizing the response data (the $Y_i$'s),
we instead require only that the fitted model $\muhat$ can only depend on
the discretized $Y_i$'s, but use the full information of the $Y_i$'s 
when computing the residuals.
\begin{itemize}
\item[(1)] \hspace{-.08in},(2) As in the naive rounded algorithm, choose $\Alg$, $\Yhat$, and $\dhat$, and compute $\muhat_y$ for each $y\in\Yhat$.
\item[(3)]
Compute the desired  quantile for the
{\em unrounded} residuals,
\[Q_y = \textnormal{Quantile}_{(1-\alpha)(1+1/n)}\left\{\big|Y_i - \muhat_y(X_i)\big|:i=1,\dots,n\right\}.\]
\item[(4)]
Finally, 
the prediction interval is given by
\begin{multline*}\PI = \left\{y'\in \R: y'\in\muhat_{\dhat(y')}(X_{n+1})\pm Q_{\dhat(y')}\right\} \\= \bigcup_{y\in\Yhat} \bigg(\dhat^{-1}(y)\cap \Big[\muhat_y(X_{n+1}) - Q_y,\muhat_y(X_{n+1})+Q_y\Big]\bigg).\end{multline*}

This prediction interval is, by construction, likely to cover the {\em unrounded} test response value, $Y_{n+1}$; it is no longer necessary to invert the rounding step.
\end{itemize}

\subsection{Coverage guarantee}

The following theorem proves the coverage properties
of the prediction intervals computed by our two discretized conformal prediction methods.\footnote{
In some settings, we may prefer a discretization function $\dhat$ that is random---for instance, if $\dhat$ rounds $y$ to the nearest value in $\Yhat$, this introduces some bias, but with randomization we can remove this bias by setting
\begin{equation}\label{eqn:random_round}
\dhat(y) = \begin{cases} y_m, &\text{ with probability }\frac{y_{m+1}-y}{y_{m+1}-y_m},\\ y_{m+1}, &\text{ with probability }\frac{y-y_m}{y_{m+1}-y_m},\end{cases}\end{equation}
where $y_m\leq y\leq y_{m+1}$ are the nearest elements to $y$ in the trial set $\Yhat$. With this construction, we 
obtain $\EEst{\dhat(y)}{y} = y$ (at least for those $y$ values that are not outside the range of the entries of $\Yhat$). 
Our main result, Theorem~\ref{thm:rounding}, can be modified to prove
the expected coverage guarantee in this setting as well, although we do not
include the details here.
}

\begin{theorem}\label{thm:rounding}
Assume exchangeability of  the training and test data points
\[(X_1,Y_1),\dots,(X_n,Y_n),(X_{n+1},Y_{n+1}).\]
Let $\Yt= \Yt\Big((X_1,Y_1),\dots,(X_{n+1},Y_{n+1})\Big)\subset\R$ be any finite set, where $\Yt$ is an exchangeable function of the $n+1$ data points. Let $\dt = \dt\Big((X_1,Y_1),\dots,(X_{n+1},Y_{n+1})\Big)$ be a discretization function, $\dt:\R\rightarrow \Yt$,
also assumed to be exchangeable in the $n+1$ data points.
Then the rounded conformal prediction interval, constructed under 
either the Conformal Prediction with Discretized Data or Conformal Prediction with a Discretized Model algorithms (presented in Section~\ref{sec:alg1} and Section~\ref{sec:alg2}, respectively), satisfies the coverage guarantee
\[\PP{Y_{n+1}\in\PI} \geq 1 - \alpha  - \PP{(\Yhat,\dhat) \neq (\Yt,\dt)}.\]
\end{theorem}

Before proving this result, we pause to note two special cases regarding the choice of the set $\Yhat$ and (randomized) discretization function $\dhat$:
\begin{itemize}
\item If $\Yhat$ and $\dhat$ are fixed (do not depend on the data), then the coverage rate is $\geq 1 - \alpha$, since we can define $\Yt = \Yhat$ and $\dt=\dhat$ always. 
\item If $\Yhat$ depends on the data only via the range of the response values, i.e.~is only a function of $\min_{i=1,\dots,n}Y_i$ and $\max_{i=1,\dots,n}Y_i$, while $\dhat$ depends only on $\Yhat$ (e.g.~$\dhat$ simply rounds any number to its nearest value in $\Yhat$, or does randomized rounding as in~\eqref{eqn:random_round}), then the coverage rate is $\geq 1 - \alpha - \frac{2}{n+1}$. This holds because, by defining $\Yt$ as the corresponding function of the range of the {\em full} data set, i.e.~of $\min_{i=1,\dots,n+1}Y_i$ and $\max_{i=1,\dots,n+1}Y_i$, we have
\[\PP{(\Yhat,\dhat)\neq (\Yt,\dt)} \leq \PP{Y_{n+1}<\min_{i=1,\dots,n}Y_i} + \PP{Y_{n+1}>\max_{i=1,\dots,n}Y_i} \leq \frac{2}{n+1}.\]
\end{itemize}

We now prove our main result.
\begin{proof}[Proof of Theorem~\ref{thm:rounding}]
Our proof closely follows the structure of the results on non-rounded conformal prediction in the earlier literature.

We begin with the naive rounded algorithm from Section~\ref{sec:alg1},
Conformal Prediction with Discretized Data.
Let \[\mut = \Alg\Big((X_1,\dt(Y_1)),\dots,(X_n,\dt(Y_n)),(X_{n+1},\dt(Y_{n+1}))\Big)\]
be the fitted model when using the complete rounded data set (i.e.~the training data as well as the test data point),
using the rounding scheme $\dt$.
Define residuals
\[\tR_i = |\dt(Y_i)-\mut(X_i)|\]
for $i=1,\dots,n+1$.
Then, by construction, we can see that $\tR_1,\dots,\tR_{n+1}$ are exchangeable, since $\dt$ and $\mut$ are both symmetric functions 
of the data $\{(X_i,Y_i):i=1,\dots,n+1\}$, and thus
\[\PP{\tR_{n+1}\leq \text{Quantile}_{1-\alpha}\big\{\tR_i:i=1,\dots,n+1\big\}} \geq 1- \alpha,\]
or equivalently,
\[\PP{\dt(Y_{n+1})\in \mut(X_{n+1}) \pm  \text{Quantile}_{(1-\alpha)(1+1/n)}\big\{\tR_i:i=1,\dots,n\big\}} \geq 1- \alpha.\]
Next, 
 on the event $(\Yhat,\dhat)=(\Yt,\dt)$, we have $\dhat(Y_i) = \dt(Y_i)$ for all $i=1,\dots,n+1$,
 and moreover, $\mut = \muhat_{\dhat(Y_{n+1})}$.
Therefore,
\begin{multline*}
\PP{\dhat(Y_{n+1})\in \PIr}\\
=
\PP{\dhat(Y_{n+1})\in \muhat_{\dhat(Y_{n+1})}(X_{n+1}) \pm  \text{Quantile}_{(1-\alpha)(1+1/n)}\big\{|\dhat(Y_i)-\muhat_{\dhat(Y_{n+1})}(X_i)|:i=1,\dots,n\big\}} \\
\geq \PP{\dt(Y_{n+1})\in \mut(X_{n+1}) \pm  \text{Quantile}_{(1-\alpha)(1+1/n)}\big\{\tR_i:i=1,\dots,n\big\}}  - \PP{(\Yhat,\dhat)\neq (\Yt,\dt)}\\
\geq 1-\alpha - \PP{(\Yhat,\dhat)\neq (\Yt,\dt)}.\end{multline*}
Finally, if $\dhat(Y_{n+1})\in\PIr$, then it holds trivially that $Y_{n+1}\in\PI = \dhat^{-1}(\PIr)$.

Next, we turn to the second algorithm, Conformal Prediction with a Discretized Model, presented
in Section~\ref{sec:alg2}. Define $\mut$ as above, and define residuals
\[\tR_i = |Y_i - \mut(X_i)|\]
for $i=1,\dots,n+1$.
As before, $\tR_1,\dots,\tR_{n+1}$ are exchangeable, and so, similarly to the calculations above, we have
\[\PP{Y_{n+1}\in \mut(X_{n+1}) \pm  \text{Quantile}_{(1-\alpha)(1+1/n)}\big\{\tR_i:i=1,\dots,n\big\}} \geq 1- \alpha.\]
Next, 
 on the event $(\Yhat,\dhat)=(\Yt,\dt)$, we have $\mut = \muhat_{\dhat(Y_{n+1})}$.
Therefore,
\begin{multline*}
\PP{Y_{n+1}\in \PI}\\
=
\PP{Y_{n+1}\in \muhat_{\dhat(Y_{n+1})}(X_{n+1}) \pm \textnormal{Quantile}_{(1-\alpha)(1+1/n)}\big\{\big|Y_i - \muhat_{\dhat(Y_{n+1})}(X_i)\big|:i=1,\dots,n\big\}} \\
\geq \PP{Y_{n+1}\in \mut(X_{n+1}) \pm  \text{Quantile}_{(1-\alpha)(1+1/n)}\big\{\tR_i:i=1,\dots,n\big\}}  - \PP{(\Yhat,\dhat)\neq (\Yt,\dt)}\\
\geq 1-\alpha - \PP{(\Yhat,\dhat)\neq (\Yt,\dt)}.\end{multline*}
\end{proof}

\subsection{Computational tradeoffs}
With our main theoretical result, Theorem~\ref{thm:rounding}, in place,
we are now able to trade off between 
the computation time of the algorithm,
and the precision of its resulting prediction intervals.
Specifically, both algorithms developed in this paper guarantee exact coverage regardless of the number of $y$ values tested (or, if $\Yhat,\dhat$ depend weakly on the data, for instance via the range of the data values, then coverage probability may decrease very slightly). 
Of course, with a smaller set $\Yhat$, the discretization will be more coarse, so the residuals will in general be larger and our resulting prediction interval will be wider. 

One interesting phenomenon that we can observe is that, if the sample size $n$ is large, then our fitted models may be highly accurate (i.e.~residuals are small) even if the added noise due to the rounding step is quite large. In other words, a low computational budget (a small set $\Yhat$ of trial values) can be offset by a large sample size.
We explore these tradeoffs empirically in the next section.

\section{Simulations}\label{sec:sims}
We now explore the effect of discretization in practice through a simulated
data experiment.\footnote{Code to reproduce
this experiment is available at \url{http://www.stat.uchicago.edu/~rina/code/CP_rounded.R}}

\paragraph{Data}
Our data is generated as
\[Y_i = \mu(X_i) + \mathcal{N}(0,\sigma^2)\]
for noise level $\sigma^2=1$, where the features are generated 
from an \iid Gaussian  model, $X_i \iidsim\mathcal{N}(0,\ident_p)$,
with dimension $p=200$. The mean function is given by
\[\mu(x) = \frac{1}{\sqrt{10}}\sum_{j=1}^{10} \left(x_j + \sign(x_j)\sqrt{|x_j|}\right),\]
so that a linear model does not fit the data exactly, but is a fairly good approximation. The sample size is $n=100$ or $n=400$.
Our model fitting algorithm $\Alg$ is given by fitting
a Lasso,
\[\muhat:x\mapsto x^\top \widehat{\beta}\text{ where }\widehat{\beta}
= \argmin_{b\in\R^p}\left\{\frac{1}{2n}\sum_{i=1}^n (Y_i - X_i^\top b)^2 + \lambda\norm{b}_1\right\},\]
for penalty parameter $\lambda = \sigma\sqrt{\frac{\log(p)}{2n}}$.
We then generate a new data point $(X_{n+1},Y_{n+1})$ from the same
distribution, and set target coverage level at $1-\alpha = 0.9$.

\begin{figure}[t]
\centering
\includegraphics[width=1.05\textwidth]{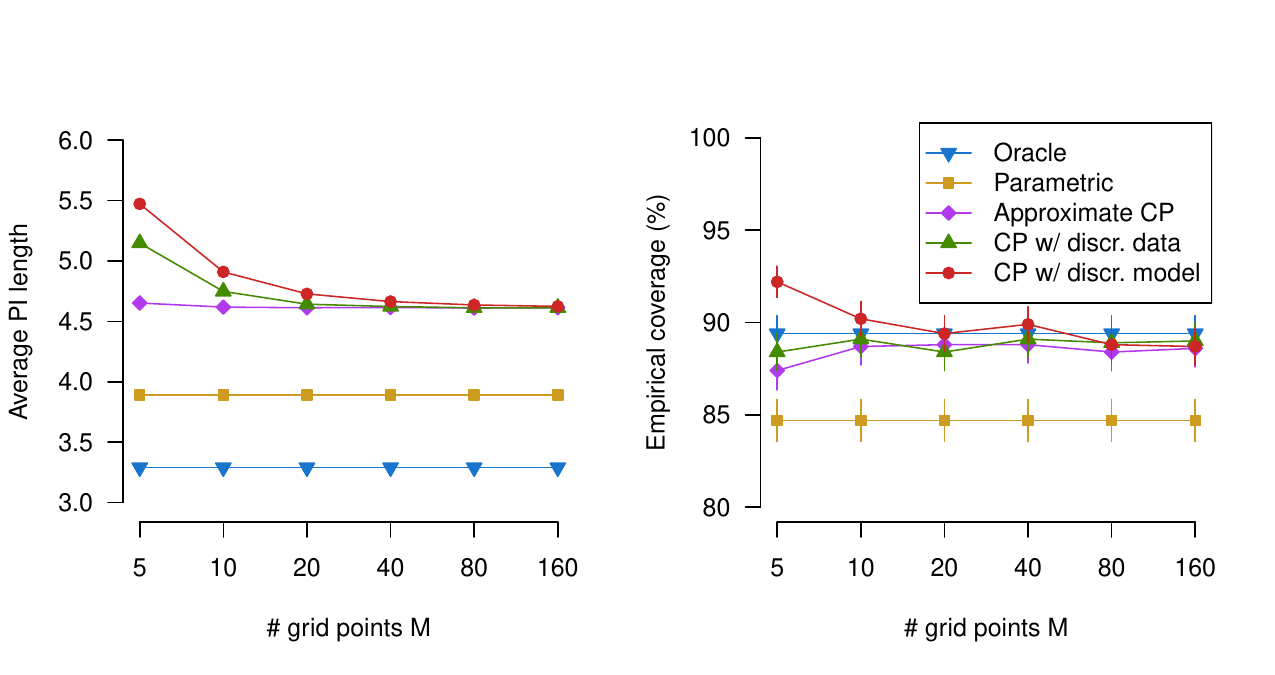}
\caption{Results of the simulated data experiment with sample size $n=100$
 (see Section~\ref{sec:sims} for details). The plots show the average prediction interval length (left) and
the empirical coverage probability (right),
along with standard error bars, for each of the five
methods considered. Results are averaged over 1000 trials.}
\label{fig:PIfig1}
\end{figure}

\paragraph{Methods}
We compare the following methods:
\begin{itemize}
\item {\bf Oracle}: using the true coefficient vector and the Gaussian distribution
of the noise, the prediction interval is given by
$\PI = \mu(X_{n+1}) \pm 1.645\sigma$.
\item {\bf Parametric}: Let $\widehat{S}=\textnormal{Support}(\widehat{\beta})$ be the support of 
the Lasso solution. If we naively compute the confidence interval for the resulting
least-squares model---that is, ignoring the fact that the feature set $\widehat{S}$ was
selected as a function of the data---we would compute a prediction interval
\[\PI = X_{n+1}^\top \widehat{\beta} \pm 1.645\sigma\sqrt{1 + X_{n+1,\widehat{S}}^\top (X_{1:n,\widehat{S}}^\top X_{1:n,\widehat{S}})^{-1}X_{n+1,\widehat{S}}}.\]
Of course, since this computation ignores the selection event, we would expect
this prediction interval to undercover.
\item {\bf Approximate Conformal Prediction} (approximated via rounding)---the informal
approximation to the conformal prediction algorithm, as presented in 
Section~\ref{sec:background_rounding}.
\item {\bf Conformal Prediction with Discretized Data (CPDD)}, as presented in Section~\ref{sec:alg1}.
\item {\bf Conformal Prediction with a Discretized Model (CPDM)}, as presented in Section~\ref{sec:alg2}.
\end{itemize}
For the rounded algorithms, we run the algorithm with grid size $M\in\{5,10,20,40,80,160\}$. The finite grid is then taken to be the set
\[\left\{y_{\min} + \frac{0.5(y_{\max}-y_{\min})}{M}, y_{\min} + \frac{1.5(y_{\max}-y_{\min})}{M}, \dots, y_{\max} - \frac{0.5(y_{\max}-y_{\min})}{M}\right\},\]
where $[y_{\min},y_{\max}]$ is the range of the observed response values $Y_1,\dots,Y_n$ in the training data. If the resulting prediction set is not an interval (which is seldom the case), we take the smallest interval containing the prediction set, for a simpler comparison.

Results are averaged over 1000 trials.

\begin{figure}[t]
\centering
\includegraphics[width=1.05\textwidth]{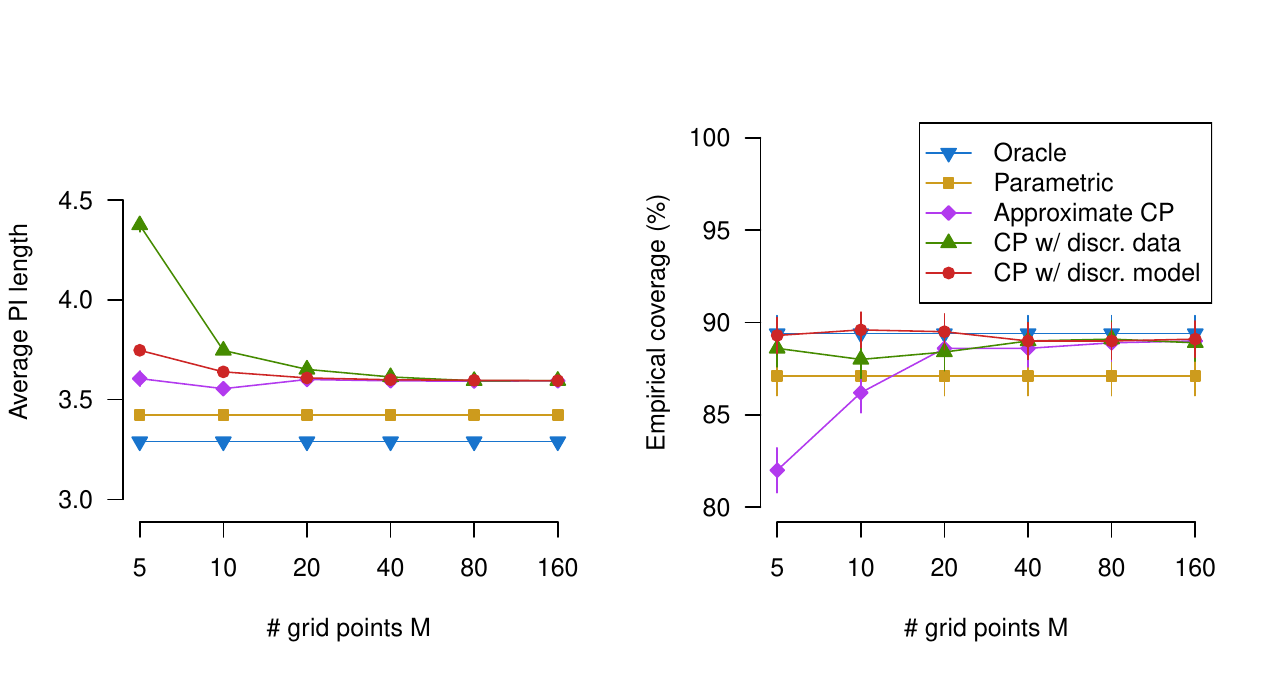}
\caption{Results of the simulated data experiment with sample size $n=400$ (details as in Figure~\ref{fig:PIfig1}).}
\label{fig:PIfig2}
\end{figure}

\paragraph{Results}
The resulting coverage probabilities and the 
average prediction interval lengths for each method
are shown in Figure~\ref{fig:PIfig1} (for sample size $n=100$) and Figure~\ref{fig:PIfig2} (for $n=400$),
across the different grid sizes $M$. (Since the oracle method and parametric method
do not use a discretized grid, reported values for these two methods are constant
across $M$.)

Examining the coverage plots first,
the oracle method has 90\% coverage as expected, and the two conformal prediction methods proposed here also show 90\% (or higher) coverage. The ``parametric'' method, by ignoring the multiple testing
problem inherent in the sparse model selection step, shows undercoverage for the smaller sample size $n=100$. At $n=400$, the selected model is more deterministic (as the signal-to-noise ratio is much stronger), so the selection event has less of an effect, and the coverage is closer
to 90\%. Finally, for the approximate conformal prediction method (i.e.~the informal rounding scheme), this method
shows the appropriate 90\% coverage 
for higher values of $M$, but can undercover when $M$ is low, particularly for the higher
sample size $n=400$. The reason is that if the grid points are spaced far apart,
while residuals tend to be fairly small, then it may be the case that none of the grid point values $y_1,\dots,y_M$
are ``plausible'' enough to get included into the prediction
interval. Therefore, this method is not reliable when the computational budget (i.e.~the number of grid points $M$) is very low.

Next, we turn to the prediction interval length results---this length represents the precision of each procedure, as an extremely wide prediction interval is not informative. The oracle method of course yields the lowest possible PI length, providing a lower bound
for the other methods. The approximate conformal prediction method (informal rounding) has somewhat lower PI length than the other rounded methods, but 
as discussed earlier, it fails to provide the guaranteed coverage rate. Comparing the two rounding algorithms proposed here, which do offer the desired coverage rate both in theory and empirically, at $n=100$ we
see similar performance, with slightly better precision (lower PI width) for the Conformal Prediction with Discretized Data (CPDD) method.
At $n=400$, however, Conformal Prediction with a Discretized Model (CPDM) gives far better performance. To understand why, recall that
for CPDD, when we discretize the data, the length of the PI will always be at least as large as the gap between two grid points; a small $M$ will therefore lead to an unfavorable lower bound on the PI length, regardless of the sample size $n$. If we  use CPDM, then coarse rounding (i.e.~a low $M$) effectively adds noise to the $y$ values, but with a sufficiently
high sample size $n$, our fitted model will be highly accurate in spite of the high effective noise level, and we can obtain low PI lengths.

\section{Summary}
In this paper, we have formalized the role of rounding and discretization
in the conformal prediction framework.
These discretized algorithms allow
conformal prediction methods to be used in practice
when computational resources are limited, while providing rigorous guarantees
that the right level of coverage will be maintained.
Our simulations demonstrate that the level of discretization can be used
to trade off between computation time and the precision of the prediction
(i.e.~the width of the prediction interval), enabling the user
to obtain meaningful guarantees at any computational budget.

\subsection*{Acknowledgements}
This work was partially supported by NSF award DMS1654076, by an Alfred P.~Sloan fellowship, and by a J.~and J.~Neubauer Faculty Development
Fellowship. The authors thank Jing Lei, Ryan Tibshirani, and Larry Wasserman for helpful discussions on the conformal prediction method, 
and Lin Gui for catching an error in an earlier draft of this paper.

\bibliographystyle{plainnat}
\bibliography{bib}

\end{document}